\documentclass[prl,aps,floats,twocolumn,showpacs,superscriptaddress,
preprintnumbers,showkeys]{revtex4-1}

\usepackage[latin1]{inputenc}                    
\usepackage{graphicx}                            
\usepackage{latexsym}                            
\usepackage{amsfonts}                            
\usepackage{amssymb}                             
\usepackage{amsmath}                             
\usepackage[mathscr]{eucal}                      
\usepackage{dcolumn}                             
\usepackage{theorem}                             
\usepackage{hyperref}                            
\graphicspath{{.}{Figs/}}                    
\newcommand{\cO}{\mathcal {O}}
\newcommand{\Dd}[1]{\overset{\leftrightarrow}{D}^{#1}}
\newcommand{\bc}{\begin{center}}
\newcommand{\ec}{\end{center}}
\newcommand{\be}{\begin{equation}}
\newcommand{\ee}{\end{equation}}
\newcommand{\bea}{\begin{eqnarray}}
\newcommand{\eea}{\end{eqnarray}}

\newcommand{\del}{\partial}
\newcommand{\gev}{{\,{\rm GeV}}}
\newcommand{\x}{\langle x\rangle}
\newcommand{\mqbar}{\overline{m}_q}

\begin{document}

\preprint{
\vbox{
\hbox{ADP-10-23/T719, DESY 10-219, Edinburgh 2010/39, LTH 893}
}}

\title[Charge Symmetry Breaking in Lattice QCD]{Charge Symmetry
  Breaking in Parton Distribution Functions from Lattice QCD}
\author{R.~Horsley}\affiliation{School of Physics and Astronomy,
  University of Edinburgh, Edinburgh EH9 3JZ, UK}
\author{Y.~Nakamura}\affiliation{Institut f\"ur Theoretische Physik,
  Universit\"at Regensburg, 93040 Regensburg, Germany}
                   \affiliation{Center for Computational Sciences, 
  University of Tsukuba, Tsukuba, Ibaraki 305-8577, Japan}
\author{D.~Pleiter}\affiliation{Deutsches Elektronen-Synchrotron DESY,
  15738 Zeuthen, Germany}
\author{P.E.L.~Rakow}\affiliation{Theoretical Physics Division, 
  Department of Mathematical Sciences, University of Liverpool, 
  Liverpool L69 3BX, UK}
\author{G.~Schierholz}\affiliation{Deutsches Elektronen-Synchrotron DESY,
  22603 Hamburg, Germany}
\author{H.~St\"uben}\affiliation{Konrad-Zuse-Zentrum f\"ur Informationstechnik Berlin, 
             14195 Berlin, Germany}
\author{A.W.~Thomas}\affiliation{CSSM, School of Physics and Chemistry, University of Adelaide,
  Adelaide SA 5005, Australia}
\author{F.~Winter}\affiliation{School of Physics and Astronomy,
  University of Edinburgh, Edinburgh EH9 3JZ, UK}
                  \affiliation{Institut f\"ur Theoretische Physik,
  Universit\"at Regensburg, 93040 Regensburg, Germany}
\author{R.D.~Young}\affiliation{CSSM, School of Physics and Chemistry,
  University of Adelaide, Adelaide SA 5005, Australia}
\author{J.M.~Zanotti}\affiliation{School of Physics and Astronomy,
  University of Edinburgh, Edinburgh EH9 3JZ, UK}
\collaboration{CSSM and QCDSF/UKQCD Collaborations} \noaffiliation

\begin{abstract}
  By determining the quark momentum fractions of the octet baryons
  from $N_f=2+1$ lattice simulations, we are able to 
  predict the degree of charge symmetry violation in the parton
  distribution functions of the nucleon. This is of importance, not
  only as a probe of our understanding of the non-perturbative
  structure of the proton but also because such a violation constrains
  the accuracy of global fits to parton distribution functions and
  hence the accuracy with which, for example, cross sections at the
  LHC can be predicted. A violation of charge symmetry may also be
  critical in cases where symmetries are used to guide the search for
  physics beyond the Standard Model.
\end{abstract}



\maketitle


Charge symmetry is related to the invariance of the QCD Hamiltonian
under rotations about the 2-axis in isospace, turning $u$ quarks to
$d$ and protons to neutrons.
Extensive studies in nuclear systems have shown that it is an
excellent symmetry~\cite{Miller:2006tv}, typically accurate to a
fraction of a percent (e.g. $m_n - m_p \sim$ 0.1\%). 
At the quark level it is, of course, very badly broken but this is
hidden by dynamical chiral symmetry breaking.
There has been extensive theoretical work on the effect of the $u-d$
mass difference on parton distribution functions (PDFs), where charge
symmetry (CS) implies~\cite{Londergan:2009kj,Londergan:1998ai}:
\begin{equation}
  \label{eq:cs}
  u^p(x,Q^2)=d^n(x,Q^2),\quad d^p(x,Q^2)=u^n(x,Q^2)\ .
\end{equation}
Within the MIT bag model, Sather~\cite{Sather:1991je} and Rodionov
{\it et al.}~\cite{Rodionov:1994cg} found that charge symmetry
violation (CSV) in the singly represented valence sector, $\delta d(x)
\equiv d^p(x) - u^n(x)$, could be as large as 5\% in the intermediate
to large range of Bjorken $x$.
Furthermore, these authors also found that $\delta u(x) \equiv u^p(x)
- d^n(x)$ was similar in magnitude but of opposite sign.

Only recently has a global analysis of PDFs allowed for CSV, with
Martin {\it et al.}~\cite{Martin:2003sk} finding a best fit that is
remarkably close to the predictions of Ref.~\cite{Rodionov:1994cg} for
both the magnitude and shape of $\delta d(x)$ and $\delta u(x)$.
Unfortunately, the errors on their result are currently too large to
be of phenomenological use but at the larger end CSV could lead to
considerable uncertainties in the predictions for some processes of
interest at the LHC.
The need for urgency in obtaining better constraints on CSV in PDFs
has recently become apparent in connection with the search for physics
beyond the Standard Model using neutrino deep-inelastic scattering.
Indeed, the level of CSV predicted in
Refs.~\cite{Sather:1991je,Rodionov:1994cg}
would reduce the $3 \sigma$ discrepancy with the Standard Model
reported by the NuTeV collaboration~\cite{Zeller:2001hh} by at least
one standard deviation~\cite{Londergan:2003pq,Bentz:2009yy}.
It was argued by Londergan and Thomas that for the second moments,
which are relevant to the NuTeV measurement, namely $\langle x \delta
d^-(x) \rangle$ and $\langle x \delta u^-(x) \rangle$ (where the
superscript minus indicates a C-odd or valence distribution function),
the results had very little model dependence~\cite{Londergan:2003ij}.
Further, future planned new-physics searches will benefit from
improved constraints on CSV, such as the parity-violating deep
inelastic scattering program at Jefferson Lab~\cite{E1207102}.

In this Letter we report the first lattice QCD determination of the
CSV arising from the $u-d$ mass difference. Our results are deduced by
studying the second moments of the parton distribution functions as we
vary the light (degenerate $u,d$) and strange quark masses in a
$N_f=2+1$ lattice simulation.
The sign and magnitude of the effect which we find are consistent both
with the estimates based on the MIT bag model~\cite{Londergan:2003ij}
and with the best fit global determination of 
Ref.~\cite{Martin:2003sk}.
However, the uncertainties in this work are considerably smaller than
those derived from the global analysis.

Because of valence quark normalisation, the first moments of $\delta
u^-(x)$ and $\delta d^-(x)$ must vanish.
\begin{table*}[tbh]
\begin{tabular}{c|c|c|c|@{\hspace{1mm}}c@{\hspace{1mm}}|@{\hspace{1mm}}c@{\hspace{1mm}}|@{\hspace{1mm}}c@{\hspace{1mm}}|@{\hspace{1mm}}c}
$\kappa_l$ & $\kappa_s$ & $m_\pi$\,[MeV]& $m_K$\,[MeV] &
${\x_u^\Sigma}/{\x_u^p}$ &
${\x_s^\Sigma}/{\x_d^p}$ &
${\x_s^\Xi}/{\x_u^p}$ &
${\x_u^\Xi}/{\x_d^p}$\\
\hline
0.12083 & 0.12104 & 460(17) & 401(15) & 1.0263(51) & 0.960(12)  & 0.993(23) & 1.044(28) \\
0.12090 & 0.12090 & 423(15) & 423(15) & 1.0        & 1.0        & 1.0       & 1.0       \\
0.12095 & 0.12080 & 395(14) & 438(16) & 0.9888(44) & 1.0344(70) & 1.010(25) & 0.985(24) \\
0.12100 & 0.12070 & 360(13) & 451(16) & 0.9670(83) & 1.059(14)  & 1.019(26) & 0.953(29) \\
0.12104 & 0.12062 & 334(12) & 463(17) & 0.9631(94) & 1.082(18)  & 1.037(29) & 0.940(30)
\end{tabular}
\caption{Pion and kaon  masses on $24^3\times 48$
  lattices with lattice spacing, $a=0.083(3)$fm
  \cite{Bietenholz:2010jr}, where the error on
  the lattice spacing has been included in the errors for $m_\pi$ and
  $m_K$. The last four columns contain our results for ratios of the
  hyperon quark momentum fractions.}
\label{tab:results}
\end{table*}
Hence the second moment (which we label $\delta q^-$) is the first
place where CSV can be visible in the valence quark distributions,
\begin{eqnarray}
  \label{eq:delu}
  \delta u^-=\int_0^1 dx\, x (u^{p-}(x)-d^{n-}(x)) &=& 
      \x^p_{u^- }- \x^n_{d^-}\,,\, \\
  \label{eq:deld}
  \delta d^-=\int_0^1 dx\, x (d^{p-}(x)-u^{n-}(x)) &=& 
      \x^p_{d^-}- \x^n_{u^-}\,.\,
\end{eqnarray}
As detailed below, these CSV momentum fractions are related to the
hyperon moments by
\begin{eqnarray}
  \label{eq:csvu}
  \delta u^-&\sim &\x_{u^-}^\Sigma - \x_{s^-}^\Xi \\
  \label{eq:csvd}
  \delta d^-&\sim &\x_{s^-}^\Sigma - \x_{u^-}^\Xi\ ,
\end{eqnarray}
in the limit where the strange and light quarks have almost equal
mass.


In the numerical calculation of these moments, 
our gauge field configurations have been generated with $N_f=2+1$
flavours of dynamical fermions, using the Symanzik improved gluon
action and nonperturbatively ${\cal O}(a)$ improved Wilson fermions
\cite{Cundy:2009yy}.
The quark masses are chosen by first finding the
SU(3)$_{\mathrm{flavour}}$-symmetric point where flavour singlet
quantities take on their physical values and 
then varying the individual quark
masses while keeping the singlet quark mass
$\mqbar=(m_u+m_d+m_s)/3=(2m_l+m_s)/3$ 
constant~\cite{Bietenholz:2010jr}.
Simulations are performed on lattice volumes of $24^3\times 48$ with
lattice spacing, $a=0.083(3)$fm.
A summary of
our dynamical configurations is given in Table~\ref{tab:results}.
More details regarding the tuning of our simulation parameters can be
found in Ref.~\cite{Bietenholz:2010jr}.

On the lattice, we compute moments of the quark distribution
functions, $q(x)$
\begin{equation}
  \label{eq:pdf}
  \langle x^{n-1}\rangle^B_q = \int^1_0 dx\,
  x^{n-1}(q^B(x)+(-1)^n\bar{q}^B(x))\ ,
\end{equation}
where $x$ is the fraction of the momentum of baryon $B$ carried by the
quarks, by calculating the matrix elements of local twist-2 operators
\begin{equation}
  \label{eq:me}
  \langle B(\vec{p})|\big[{\cal
    O}_q^{\{\mu_1\ldots\mu_n\}}-\mathrm{Tr}\big]|B(\vec{p})\rangle =
  2\langle x^{n-1}\rangle^B_q [p^{\mu_1}\cdots p^{\mu_n} - 
  \mathrm{Tr}]\ ,
\end{equation}
where
$  {\cal O}_q^{\mu_1\ldots\mu_n}=i^{n-1}
  \bar{q}\gamma^{\mu_1}\Dd{\mu_2}\cdots\Dd{\mu_n}q\ .
$

In this paper we consider only the quark-line connected contributions
to the second $(n=2)$ moment, $\x_q$, which means we only include the
part of $\bar{q}^B$ coming from quark-line connected backward
moving quarks, the so-called Z-graphs.
While the contributions from disconnected insertions are expected
to be small, in the following analysis we will focus on differences of
baryons and so these contributions will cancel in the
SU(3)$_{\mathrm{flavour}}$ limit and should be negligible for small
expansions around this limit, as considered here.

We use the standard local operator
$    {\cal O}_q^{\x}={\cal O}_q^{44}-1/3
({\cal O}_q^{11} + {\cal O}_q^{22} + {\cal O}_q^{33})
$\, .
The matrix element in Eq.~(\ref{eq:me}) is obtained on the lattice by
considering the ratio:
\begin{equation}
  \label{eq:rat}
  R(t,\tau,\vec p)=\frac{C_{\mathrm{3pt}}(t,\tau,\vec
    p)}{C_{\mathrm{2pt}}(t,\vec p)} 
     = -\frac{E^2_{\vec{p}}+\frac{1}{3}\vec{p}\,^2}{E_{\vec p}} 
     \x\ ,
\end{equation}
where $C_{\mathrm{2pt}}$ and $C_{\mathrm{3pt}}$ are lattice two and
three-point functions, respectively, with total momentum, $\vec p$,
(in our simulation we consider only $\vec{p}=0$).
The operator ${\cal O}_q^{\x}$ 
is inserted into the three-point function,
$C_{\mathrm{3pt}}(t,\tau,\vec p)$ at time, $\tau$, between the baryon
source located at time, $t=0$, and sink at time, $t$.

\begin{figure}[tb]
\bc
\vspace*{-2mm}
\includegraphics[width=0.49\textwidth]{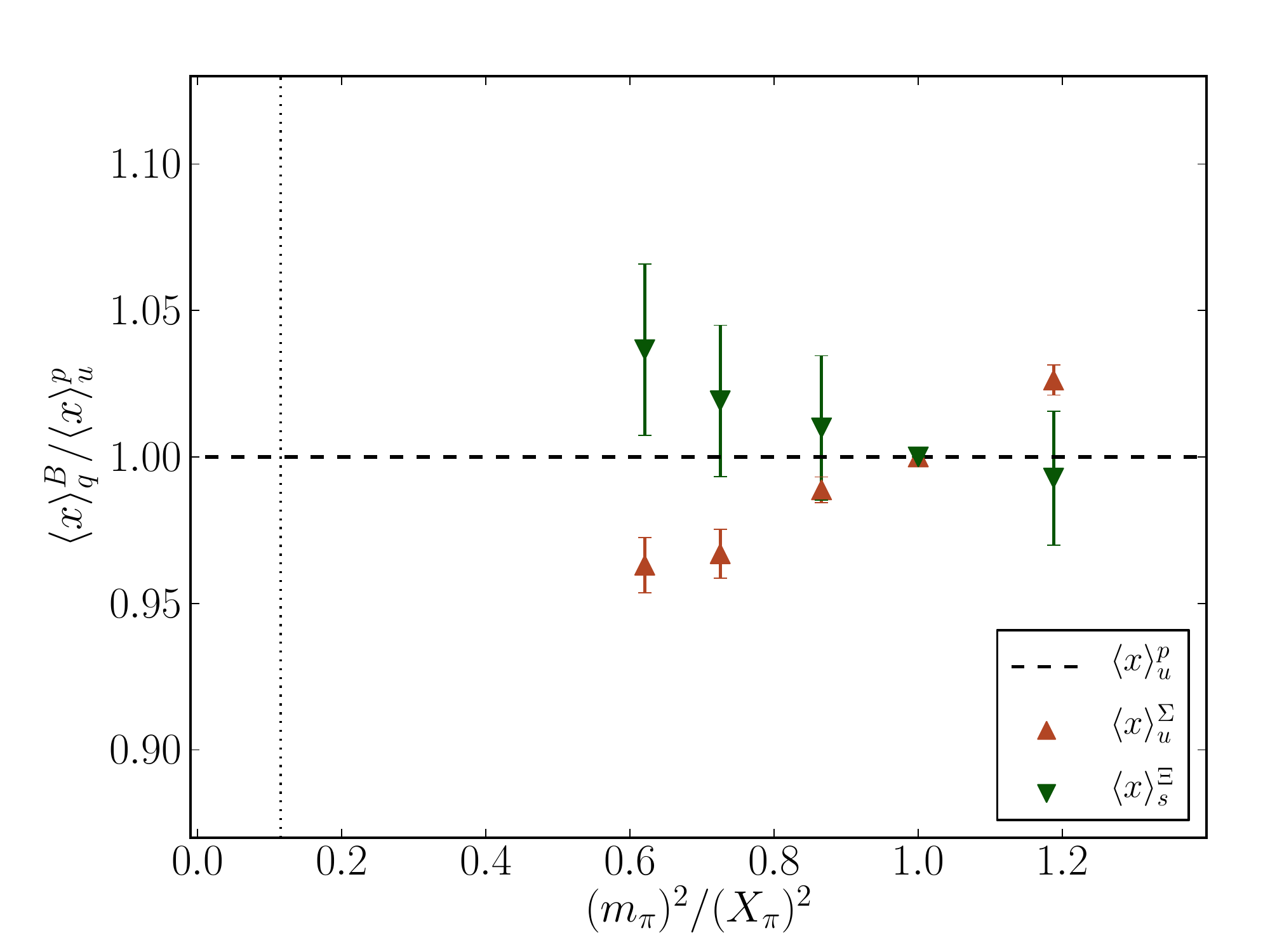}
\caption{Ratio of doubly represented quark momentum fractions,
  $\x_u^\Sigma/\x_u^p$ and $\x_s^\Xi/\x_u^p$ as a function of
  $m_\pi^2/X_\pi^2$, where we have determined $X_\pi$ from the masses
  in Tab.~\ref{tab:results}.}
\label{fig:xu}
\ec
\end{figure}
\begin{figure}[tb]
\bc
\vspace*{-2mm}
\includegraphics[width=0.49\textwidth]{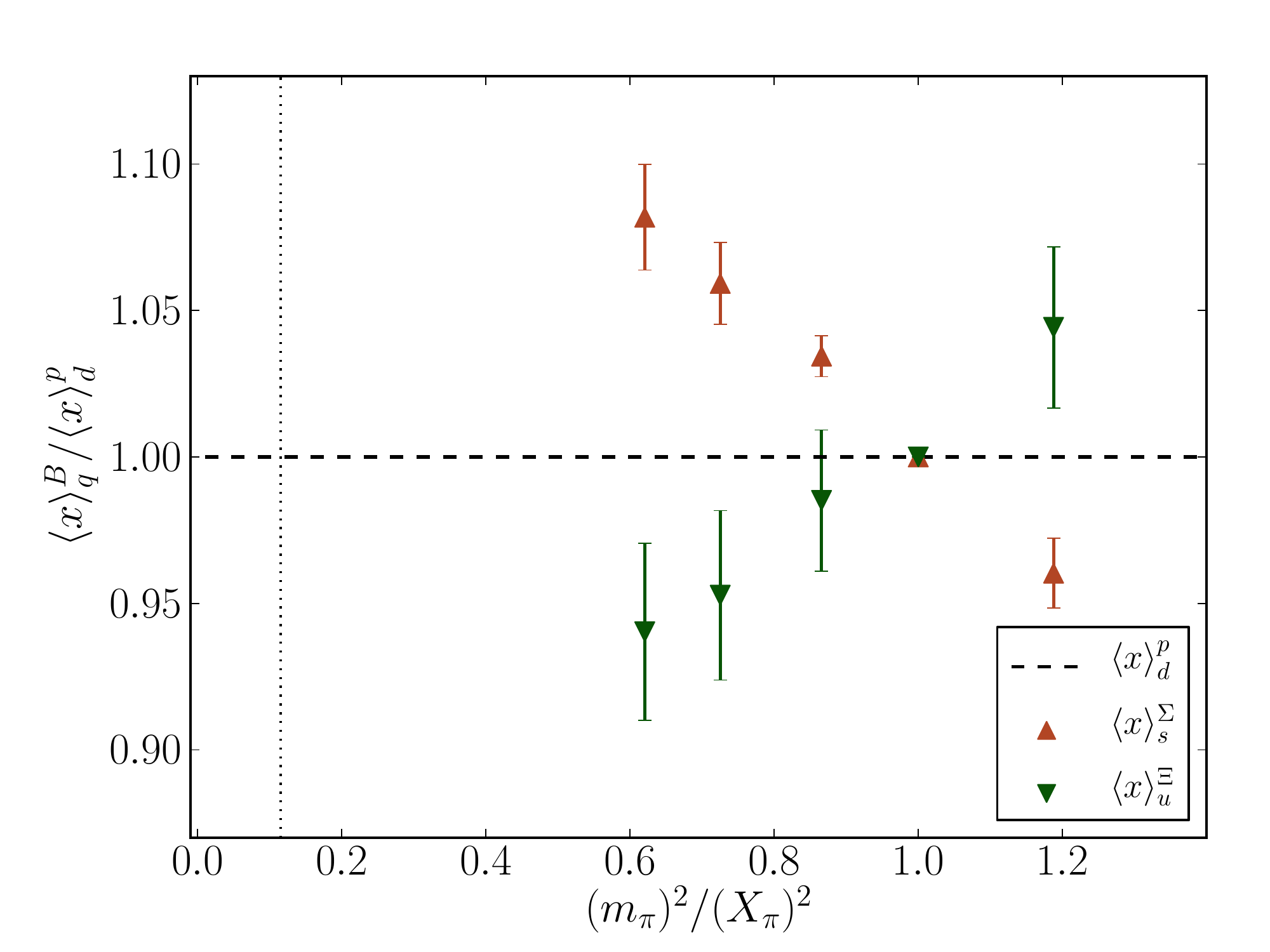}
\caption{Ratio of singly represented quark momentum fractions,
  $\x_s^\Sigma/\x_d^p$ and $\x_u^\Xi/\x_d^p$ as a function of
  $m_\pi^2/X_\pi^2$, where we have determined $X_\pi$ from the masses
  in Tab.~\ref{tab:results}.}
\label{fig:xd}
\ec
\end{figure}

The operators used for determining the quark momentum fractions need
to be renormalised, preferably using a nonperturbative method such as
RI$^\prime$-MOM \cite{Martinelli:1994ty,Gockeler:1998ye,Gockeler:2010yr}.
Here, however, we will only present results for ratios of quark
momentum fractions so that the renormalisation constants cancel and
hence our results are scale and scheme independent.
In Fig.~\ref{fig:xu} we present results for the ratio of the
$u(s)$-quark momentum fraction of the $\Sigma(\Xi)$ baryon to the
momentum fraction of the $u$ in the proton. They are also given in
Table~\ref{tab:results}, as a function of $m_\pi^2$, normalised with the
centre-of-mass of the pseudoscalar meson octet,
$X_\pi=\sqrt{(2m_K^2+m_\pi^2)/3}=411$~MeV.
Here we see the strong effect of the decrease (increase) in the light
(strange) quark momentum fractions as we approach the physical point.
In particular, we see that the heavier strange quark in the $\Xi^0$
carries a larger momentum fraction than the up quark in the proton.
We also notice that the up quark in the $\Sigma^+$ has a smaller
momentum fraction than the up quark in the proton.
This is a purely environmental effect since the only difference
between these two measurements is the mass of the spectator quark ($s$
in $\Sigma^+$, $d$ in $p$).
This implies that the momentum fraction of the strange quark
in the $\Sigma$ should be larger than that of the down quark in the
proton, which is exactly what we see in Fig.~\ref{fig:xd}.

To infer the level of CSV relevant to the nucleon, we only need to
consider a small expansion about the
SU(3)$_{\mathrm{flavour}}$ symmetric point, for which linear flavour
expansions prove to work extremely well \cite{Bietenholz:2010jr}.
For instance, we can write
\begin{equation}
\delta u= m_\delta\left(-\frac{\del\x_u^p}{\del m_u}
+\frac{\del\x_u^p}{\del m_d}\right)+\cO(m_\delta^2)\,,
\label{eq:du}
\end{equation}
where $m_\delta\equiv (m_d-m_u)$ and we have already made use of
charge symmetry by equating 
$\del\x_d^n/\del m_d=\del\x_u^p/\del m_u$ 
and $\del\x_d^n/\del m_u=\del\x_u^p/\del m_d$. A similar expression holds
for $\delta d$.

Near the SU(3)$_{\mathrm{flavour}}$ symmetric point, we note that the
up quark in the proton is equivalent to an up quark in a $\Sigma^+$ or
a strange quark in a $\Xi^0$, which we describe collectively as the
``doubly-represented'' quark \cite{Leinweber:1995ie}.

The local derivatives required for $\delta u$ can be obtained by
varying the masses of the up and down quarks independently. Within the
present calculation, we note that the difference $\x_s^{\Xi}-\x_u^p$ measures precisely the
variation of the doubly-represented quark matrix element with respect
to the doubly-represented quark mass (while holding the
singly-represented quark mass fixed). Similar variations allow us to
evaluate the other required derivatives, where we write
\begin{eqnarray}
\frac{\del \x_u^p}{\del m_u}&\simeq&
\frac{\x_s^{\Xi^0}-\x_u^p}{m_s-m_l}\,,\ 
\frac{\del \x_u^p}{\del m_d}\simeq \frac{\x_u^{\Sigma^+}-\x_u^p}{m_s-m_l}\,,\\
\frac{\del \x_d^p}{\del m_u}&\simeq&
\frac{\x_u^{\Xi^0}-\x_d^p}{m_s-m_l}\,,\ 
\frac{\del \x_d^p}{\del m_d}\simeq \frac{\x_s^{\Sigma^+}-\x_d^p}{m_s-m_l}\,.
\end{eqnarray}
With these expressions and Eq.~(\ref{eq:du}), we obtain the relevant
combinations for our determination of CSV in the nucleon
\begin{equation}
\delta u=m_\delta\frac{\x_u^{\Sigma^+}-\x_s^{\Xi^0}}{m_s-m_l}\,,\ 
\delta d=m_\delta\frac{\x_s^{\Sigma^+}-\x_u^{\Xi^0}}{m_s-m_l}\,.
\end{equation}

\begin{figure}
\bc
\vspace*{-2mm}
\includegraphics[width=0.49\textwidth]{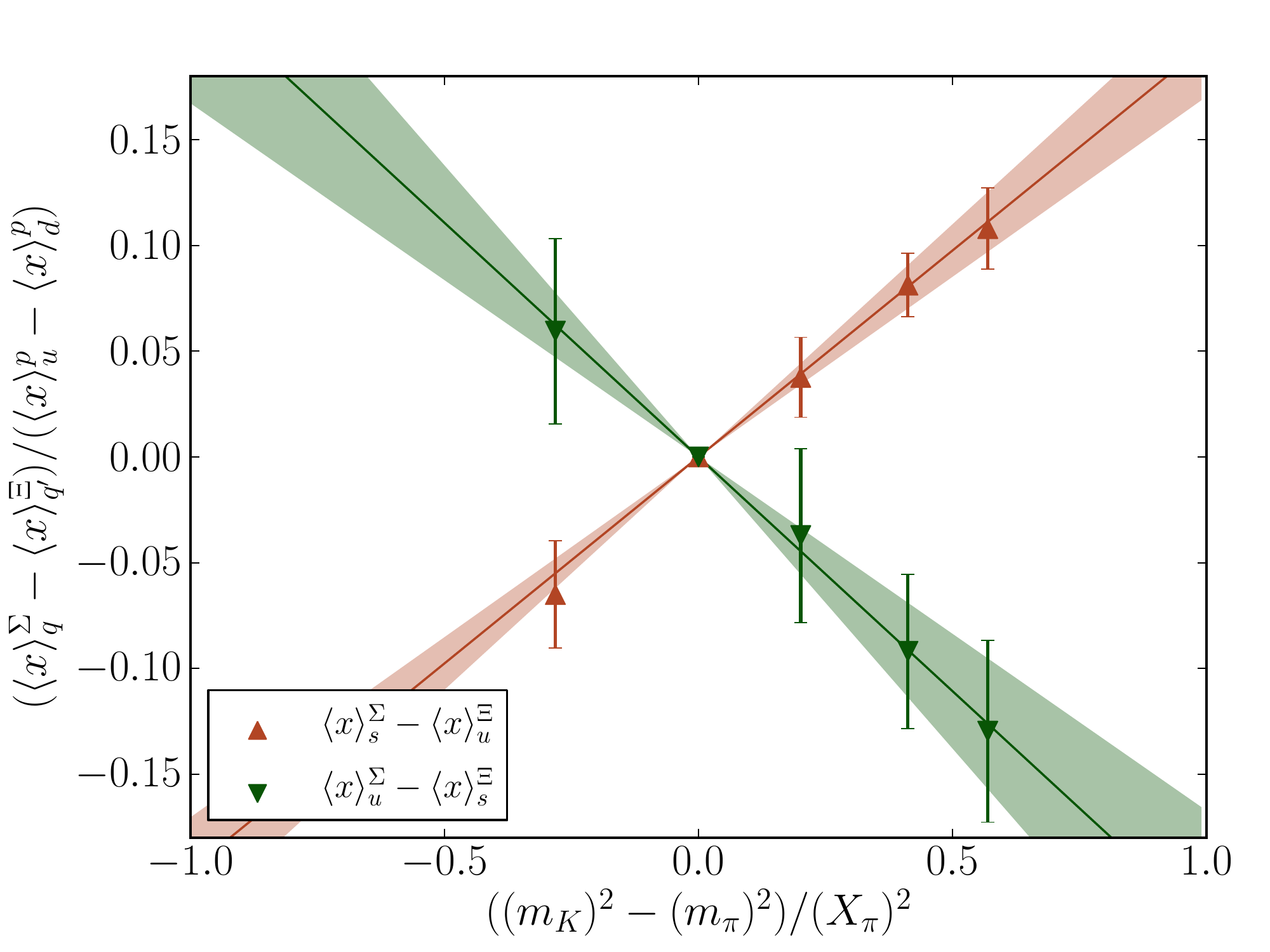}
\caption{The difference between the doubly and singly represented
  quarks in the $\Sigma$ and $\Xi$ as a function of the strange/light
  quark mass difference. We deduce $\delta u$ and $\delta d$,
  respectively, from the slopes of these curves
  (c.f. Eqs.~(\ref{eq:deltau}) and (\ref{eq:deltad})).}
\label{fig:xud}
\ec
\end{figure}

By invoking the Gell-Mann--Oakes--Renner relation and normalising to
the total nucleon isovector quark momentum fraction, we write
\begin{eqnarray}
\frac{\delta u}{\x_{u-d}^p}
&=&\frac{m_\delta}{\mqbar}
\frac{(\x_u^{\Sigma^+}-\x_s^{\Xi^0})/\x_{u-d}^p}{(m_K^2-m_\pi^2)/X_{\pi}^2}\,,\\
\frac{\delta d}{\x_{u-d}^p}
&=&\frac{m_\delta}{\mqbar}
\frac{(\x_s^{\Sigma^+}-\x_u^{\Xi^0})/\x_{u-d}^p}{(m_K^2-m_\pi^2)/X_{\pi}^2}\,.
\end{eqnarray}
Written in this way, the fractional CSV terms are just the slopes of
the curves shown in Fig.~\ref{fig:xud} (evaluated at the symmetry
point) multiplied by the ratio $m_\delta/\mqbar$. By fitting the
slopes, we obtain
\begin{eqnarray}
\frac{\delta u}{\x_{u-d}^p}&=&\frac{m_\delta}{\mqbar}(-0.221\pm 0.054)
\label{eq:deltau} \\
\frac{\delta d}{\x_{u-d}^p}&=&\frac{m_\delta}{\mqbar}(0.195\pm 0.025)
\label{eq:deltad}
\end{eqnarray}

Chiral perturbation theory yields the quark mass ratio
$m_\delta/\mqbar=0.066(7)$ \cite{Leutwyler:1996qg}
and the isovector momentum fraction is experimentally determined 
to be $\x_{u-d}^p\simeq 0.158$ at $4\gev^2$. Substituting these values 
into Eqs.~(\ref{eq:deltau}) and (\ref{eq:deltad}) yields
the first lattice QCD estimates of the CSV momentum fractions
\begin{eqnarray}
\delta u=-0.0023(6),
\quad 
\delta d=0.0020(3).
\end{eqnarray}
The first observation we make is that these results are roughly equal
in magnitude and have opposite sign. These values are slightly larger
than, but within errors in agreement with, the phenomenological
predictions of ~\cite{Rodionov:1994cg,Londergan:2003pq}, where within
the MIT bag model (at a scale $Q^2\simeq 4\gev^2$) they found $\delta
u^-=-0.0014$ and $\delta d^-=0.0015$. They are also consistent with
the best-fit values of the phenomenological analysis of
MRST~\cite{Martin:2003sk}, $\delta u^-=-\delta
d^-=-0.002^{+0.009}_{-0.006}$ (90\% CL).

While our work provides the first nonperturbative QCD result to give a
clear indication of the sign and magnitude of the CSV in these
moments, we point out that it is based on lattice simulations using a
single volume and lattice spacing.
To achieve a precise quantitative determination will require a
detailed study of the finite-volume and discretisation effects which
we plan to address by extending these calculations to larger volumes
and a second lattice spacing.
Additionally, it is well known that lattice results for the second
moment of the iso-vector nucleon PDFs, $\x_{u-d}$, do not agree well
with experiment (see e.g. \cite{Hagler:2009ni}).
Based on chiral perturbation theory it is expected that finite size
effects and chiral corrections are potentially large
\cite{Detmold:2001jb,Detmold:2003rq,Detmold:2005pt,Dorati:2007bk}, but
this has so far not been confirmed by lattice calculations.
This discrepancy may also be due to a mismatch of lattice nucleon
matrix elements and perturbative Wilson coefficients.
However, what concerns us here are ratios of moments of PDFs, in which
such effects cancel out. 
For example, we find $\x_u^p/\x_d^p \approx 2.3$ in good
agreement with $\x_{u^-}^p/\x_{d^-}^p = 2.40(6)$ found
in~\cite{Blumlein:2006be}.
We are also encouraged that lattice results for the ratio
$\x_{(u-d)}/\x_{(\Delta u - \Delta d)}$ agree well with experiment
\cite{Aoki:2010xg}.
Lastly, we have estimated the CSV associated only with the $u-d$ mass
difference.
It is important to also find a method to investigate the CSV induced
by electromagnetic effects which is expected
\cite{Martin:2003sk,Gluck:1998xa} to be of a similar size.
The determination of this effect is, however, a separate calculation
which will have no impact on our result.

%


In summary, we have performed the first lattice determinations of the quark
momentum fractions of the hyperons, $\Sigma$ and $\Xi$ in $N_f=2+1$
lattice QCD.
By examining the SU(3)$_{\mathrm{flavour}}$-breaking effects in these
momentum fractions, we are able to extract the first QCD 
determination of the size and
sign of charge-symmetry violations in the parton distribution functions
in the nucleon, $\delta u$ and $\delta d$.
Although our lattice calculations are restricted to the second $(n=2)$
moment of the C-even quark distributions, our results for $\delta
u=-0.0023(6)$, $\delta d=0.0020(3)$ are in excellent agreement with
earlier phenomenological
calculations~\cite{Rodionov:1994cg,Londergan:2003pq}.

%
\section*{Acknowledgements}
%

The numerical calculations have been performed on the apeNEXT at
NIC/DESY (Zeuthen, Germany), the IBM BlueGeneL at EPCC (Edinburgh,
UK), the BlueGeneP (JuGene) and the Nehalem Cluster (JuRoPa) at NIC
(J\"ulich, Germany), and the SGI ICE 8200 at HLRN (Berlin-Hannover,
Germany).
We have made use of the Chroma software suite \cite{Edwards:2004sx}.%
This work has been supported in part by the DFG (SFB/TR 55, Hadron
Physics from Lattice QCD) and the EU under grants 238353
(ITN STRONGnet) and 227431 (HadronPhysics2).
JZ is supported by STFC under contract number ST/F009658/1.
This work was also supported by the University of Adelaide and the Australian
Research Council through an Australian Laureate Fellowship (AWT).
%
%

\bibliography{CSV_References}


\end{document}